\documentclass[11pt]{article}
\usepackage{amsmath,amsfonts,amsthm,amssymb}

\newcommand{\version}{Dec. 2, 2004}

\font\notefont=cmsl8 

\newcommand\beq{\begin{equation}}
\newcommand\eeq{\end{equation}}

\newcommand\N{{\mathbb N}}
\newcommand\C{{\mathbb C}}

\newcommand\R{{\mathbb R}}
\newcommand\eps{\varepsilon}
\newcommand\half{\mbox{$\frac 12$}}
\renewcommand\rho\varrho

\newcommand\Tr{{\rm Tr}}

\newcommand{\U}{{\cal U}}                                                       
\newcommand\Hh{{\cal H}}

\newcommand\Bb{{\cal B}}
\newcommand\Kk{{\cal K}}
\newcommand\id{{\mathbb I}}
\newcommand{\KK}{{K^{\alpha *}}}
\newcommand{\K}{{K^\alpha}}
\newcommand{\Oo}{{\cal O}}
\newcommand{\nl}{\, \mbox{$< \!\!\!\! \vert$}\,\,\, }
   
\newtheorem{thm}{Theorem}

\newtheorem{cor}{Corollary}
\theoremstyle{definition}

\begin{document}

\title{A Stronger Subadditivity of Entropy}

\author{\vspace{5pt} Elliott H.~Lieb$^{1}$ and Robert Seiringer$^{2}$\\
\vspace{-3pt}\small{Departments of Physics$^{1,2}$ and Mathematics$^{1}$, Jadwin Hall,}\\ \vspace{-3pt}
\small{Princeton University, P.~O.~Box 708, Princeton, New Jersey
  08544, USA.}\\ 
\small{Email: \{\texttt{lieb, rseiring}\}\texttt{@princeton.edu}}}
\date{\small \version}
\maketitle

\renewcommand{\thefootnote}{$1$}
\footnotetext{Work partially
supported by U.S. National Science Foundation
grant PHY 01 39984.}
\renewcommand{\thefootnote}{$2$}
\footnotetext{Work partially
supported by U.S. National Science Foundation
grant PHY 03 53181, and by an A.P. Sloan Fellowship\\
\copyright\, 2004 by the authors. This paper may be reproduced, in its
entirety, for non-commercial purposes.}

\begin{abstract}
  The strong subadditivity of entropy plays a key role in several
  areas of physics and mathematics. It states that the entropy
  $S[\rho]= -\Tr\,\big( \rho \ln \rho\big) $ of a density matrix
  $\rho_{123}$ on the product of three Hilbert spaces satisfies
  $S[\rho_{123}] - S[\rho_{23}] \leq S[\rho_{12}]- S[\rho_2]$.  We
  strengthen this to $S[\rho_{123}] - S[\rho_{12}] \leq \sum_\alpha
  n^\alpha \big( S[\rho_{23}^\alpha ] - S[\rho_2^\alpha ]\big)$, where
  the $n^\alpha $ are weights and the $\rho_{23}^\alpha$ are
  partitions of $\rho_{23}$.  Correspondingly, there is a
  strengthening of the theorem that the map $A\mapsto \Tr \exp[L + \ln
  A] $ is concave. As applications we prove some monotonicity and
  convexity properties of the Wehrl entropy and entropy inequalities
  for quantum gases.
\end{abstract}
\bigskip

\section{Introduction}

The strong subadditivity of entropy (SSA), whose proof in the
non-com\-mu\-ta\-tive case was given by Lieb and Ruskai
\cite{LR,LBull}, is one of the main ingredients in various fields of
mathematics and physics in which the von Neumann/Shannon entropy plays
a role. Over the years other proofs have appeared
\cite{NaTh,petz,LCarlen}. SSA is an inequality among various entropies
that can be formed from one density matrix on the product of three
Hilbert spaces $\Hh_{123} = \Hh_1\otimes\Hh_2\otimes \Hh_3$ and states
that
\begin{equation}\label{ssa}
S[\rho_{123}] - S[\rho_{12}] \leq S[\rho_{23}]- S[\rho_2] \ .  
\end{equation}
Here, $\rho_{123} $ is a density matrix (i.e., a positive
semi-definite operator whose trace is 1) on the tensor product space 
$\Hh_{123} $, and $\rho_{12} $ is the reduced density matrix on
$\Hh_{12} = \Hh_1\otimes \Hh_2$, formed by taking the partial trace of
$\rho_{123} $ over $\Hh_3$ (i.e., $\rho_{12}= \Tr_{\Hh_3}\rho_{123}$),
and so forth. The entropy $S[\rho]$ on a Hilbert space $\Hh$ is given
by the von Neumann/Shannon formula
\begin{equation} 
S[\rho] = - \Tr_{\Hh}\left( \rho \ln \rho \right).  
\end{equation}
(Henceforth, the Hilbert space notation $\Hh$ on the trace, $\Tr\, $,
will be omitted if it is not needed, or we may simply write $\Tr_1$ to
denote $\Tr_{\Hh_1}\,$, etc.; likewise, $S_{123}$ will denote
$S[\rho_{123}]\,$, etc., when the meaning is clear.)

Inequality (\ref{ssa}) appears to be straightforward enough that it
seems unlikely that it can be improved, i.e., that one can insert a
quantity between the left and right sides that preserves the
inequality.  That, however, is what we do in this paper (cf.  Eq.
(\ref{intermed})). Admittedly, our theorems can be derived from SSA
(or, equivalently \cite{lindblad}, from the monotonicity of relative
entropy under completely positive trace preserving (CPT) maps) and thus,
when viewed from a sufficiently remote perspective, there is little
new here. From the point of view of applications and of understanding
the potential of SSA, however, our results and proof techniques may
merit attention, especially our applications to the theory of quantum
gases in Corollary~\ref{Cf} of Section~\ref{Cfsect}.

Inequality (\ref{ssa}) is written in a slightly unusual way. Instead
of the usual $S_{123} +S_2 \leq S_{12} +S_{23}\,$, (\ref{ssa})
displays the inequality as the decrease of the conditional entropy
$S_{23} - S_2$ when 2 is replaced by 12, i.e., information about the
state on $\Hh_1$ is added. Our focus will be on the conditional
entropy.

In Section~\ref{applications} we give some examples of the utility of
the improved version of inequality (\ref{ssa}), Eq. (\ref{teqmain}).
As one example, we show that the \lq mutual information\rq\ $S_{1} +
S_2 - S_{12}$ is decreased if the density matrix is replaced by
Wehrl's corresponding classical phase-space function (whose definition
will be recalled later). Wehrl had shown \cite{wehrl} that his entropy
is always greater than the true entropy, but the monotonicity of the
difference $S_{12} -S_1- S_2$ is new. This is a special case of
Corollary~\ref{C2} below. We also show that the difference between the
Wehrl and the true entropy is a convex function of the density matrix.

Originally, we had proved the monotonicity of $S_{12}-S_1$, and we are
grateful to M.B. Ruskai for suggesting the stronger version to us; her
argument, which uses the theory of CPT maps, is briefly sketched in
Appendix~\ref{ruskapp} \cite{rus}. We also acknowledge other helpful
correspondence about this paper.

In another direction, it will be recalled that one of the ways to
prove SSA is by means of the theorem \cite{L} (for one Hilbert space)
that the map
\begin{equation}\label{concavi}
A \mapsto \Tr\, \exp \big( L + \ln A \big) 
\end{equation} 
for positive definite operators $A$ is concave for each fixed
self-adjoint $L$.  This, too, will be improved, and its improvement
will lead to the improved version of SSA.

Our main result is the following.

\begin{thm}[Stronger Subadditivity]\label{Tmain}
  Let $\Hh_i$, $i=1,2,3$, be separable Hilbert spaces, and let $\rho_{123}$ be a
  density matrix on $\Hh_{123}=\Hh_1\otimes\Hh_2\otimes \Hh_3$ with
  finite entropy.  Let $\Omega$ be a measure space, with elements
  labeled by $\alpha$, and let $d\mu(\alpha)$ be a measure on
  $\Omega$.  Let $K^\alpha$ be bounded operators on
  $\Hh_{12}=\Hh_1\otimes \Hh_2$ that are weakly measurable, and
  satisfy (with $\KK$ the adjoint of $\K$)
\begin{equation} \label{enmain}
\int_\Omega d\mu(\alpha)\,  K^{\alpha *}  K^{\alpha } =\id_{\Hh_{12}} \ . 
\end{equation}
With the usual notational abuse $\K \leftrightarrow \K\otimes \id_{\Hh_3}$, let
\begin{equation}
n^\alpha = \Tr_{123}\,  K^{\alpha } \rho_{123}K^{\alpha * }
\end{equation}
and, in case $n^\alpha>0$, let 
\begin{align}\label{rho1main}
\rho_{23}^\alpha &= \Tr_1 \, K^{\alpha } \rho_{123}K^{\alpha*} / n^\alpha \ , \\
\rho_{2}^\alpha & = \Tr_{3}\, \rho_{23}^\alpha 
= \Tr_{13} \, K^{\alpha } \rho_{123}K^{\alpha*} / n^\alpha \ . \label{rho3main}
\end{align}
Then
\begin{equation}\label{teqmain}
S[\rho_{123}] - S[\rho_{12}] \leq    
\int_\Omega d\mu(\alpha) \, n^\alpha \big(  S[\rho_{23}^\alpha]- S[\rho_2^\alpha] \big).
\end{equation}
\end{thm}

\newpage
\noindent {\bf Remarks.} 
\begin{enumerate}
  
\item We recall that weakly measurable means that $\langle \phi |
  K^{\alpha}| \psi\rangle$ is measurable for any vectors
  $|\phi\rangle$ and $|\psi\rangle$ in the Hilbert space. (This is
  implied, via polarization, by the seemingly weaker condition of
  measurability for all $|\phi\rangle=|\psi\rangle$.) The integrals
  then have to be interpreted in the weak sense, e.g., (\ref{enmain})
  means that $\int_\Omega d\mu(\alpha)\, \langle \phi | \KK \K|
  \psi\rangle = \langle\phi|\psi\rangle$ for all $|\phi\rangle$ and
  $|\psi\rangle$.
  
\item Because of cyclicity of the trace, $n^\alpha = \Tr\, \KK\K
  \rho_{123}$, and hence (\ref{enmain}) implies that $\int_\Omega
  d\mu(\alpha)\, n^\alpha = \Tr\, \rho_{123}= 1$.

\item Both sides of the inequality (\ref{teqmain}) are homogeneous of order
1 in $\rho_{123}$. Hence this inequality holds also without the
normalization condition $\Tr\, \rho_{123}=1$, i.e., it holds for all
positive trace class operators.

\item It is no restriction to assume the $\K$ to be bounded. Because
  of (\ref{enmain}), they must be bounded almost everywhere, and hence
  one can absorb their norm into the measure $d\mu(\alpha)$.

\item In case all the $K^\alpha$ act non-trivially only on $\Hh_1$
  (i.e., $K^\alpha = k^\alpha \otimes \id_2$ and $\int_\Omega
  d\mu(\alpha)\, k^{\alpha *} k^{\alpha } =\id_{\Hh_{1}}$) we have
  that $\int_\Omega d\mu(\alpha)\, n^\alpha\rho_{23}^\alpha=
  \rho_{23}$.  Since the map $\rho_{23} \mapsto S[\rho_{23}]-
  S[\rho_2]$ is concave, as shown in \cite{LR}, the right side of
  (\ref{teqmain}) is bounded above by $S[\rho_{23}]- S[\rho_2]$ in this
  special case.  Theorem~\ref{Tmain} is, therefore, stronger than the
  usual strong subadditivity of entropy because we have
\begin{equation}\label{intermed}
S[\rho_{123}] - S[\rho_{12}] \leq    
\int_\Omega d\mu(\alpha) \, n^\alpha \big(  S[\rho_{23}^\alpha]- S[\rho_2^\alpha] \big)
\leq S[\rho_{23}] - S[\rho_{2}] 
\end{equation}
in this case.

\item Everything remains true in the classical case as well. That is,
  $\Hh_i$ is replaced by a measure space, $\rho$ is replaced by a
  measurable function and the trace is replaced by an integral. The
  $K^\alpha$ are then functions on the measure space. Note that in the
  limit that $\KK\K$ is just a $\delta$-function supported at a point
  in the measure space, labeled by $\alpha$, Inequality
  (\ref{teqmain}) is actually an {\it equality} in the classical case.
\end{enumerate}

\subsection{The special case of matrices and sums}

To keep things simple we shall first deal with the finite dimensional
case, when $\Hh_i = \C^{n_i}$ for finite $n_i$, and with the case
where the integral in (\ref{enmain}) is just a finite sum. In this
special case, Theorem~\ref{Tmain} is then just Theorem~\ref{T1} below.
We will first prove Theorem~\ref{T1}. The extension to the case of a
general measure space in (\ref{enmain}) is given in
Appendix~\ref{intapp}, and the extension to the infinite dimensional
case is given in Appendix~\ref{infapp}.  

\begin{thm}[Stronger Subadditivity, Matrix Case]\label{T1}
  Let $\rho_{123}$ be a density matrix on a finite dimensional Hilbert
  space $\Hh_{123}= \Hh_1\otimes\Hh_2\otimes \Hh_3$.  Let $K^\alpha$,
  $1\leq \alpha\leq M$, be a finite set of operators on
  $\Hh_{12}=\Hh_1\otimes \Hh_2$, that satisfy
\begin{equation} \label{en}
\sum_\alpha K^{\alpha *}  K^{\alpha } =\id_{\Hh_{12}} \ . 
\end{equation}
With the usual notational abuse $\K \leftrightarrow \K\otimes \id_{\Hh_3}$, let
\begin{equation}\label{rho1a}
n^\alpha = \Tr_{123}\,  K^{\alpha } \rho_{123}K^{\alpha * }
\end{equation}
and, in case $n^\alpha>0$, let
\begin{align}\label{rho1}
\rho_{23}^\alpha &= \Tr_1 \, K^{\alpha } \rho_{123}K^{\alpha*} / n^\alpha \ , \\
\rho_{2}^\alpha & = \Tr_{3}\, \rho_{23}^\alpha 
= \Tr_{13} \, K^{\alpha } \rho_{123}K^{\alpha*} / n^\alpha \ . \label{rho3}
\end{align}
Then
\begin{equation}\label{teq}
S[\rho_{123}] - S[\rho_{12}] \leq    
\sum_\alpha n^\alpha \big(  S[\rho_{23}^\alpha]- S[\rho_2^\alpha] \big).
\end{equation}
\end{thm}

We will give two independent proofs of Theorem~\ref{T1} in the next
section. The first one uses Theorem~\ref{L1} below, which states the
generalization of the concavity in (\ref{concavi}) mentioned above.
The second proof uses the theory of completely positive maps.

\begin{thm}[A concave map]\label{L1} 
Let $L$  be a self-adjoint operator on a finite dimensional Hilbert space $\Hh$. 
For $1\leq \alpha\leq M$, 
let $A^\alpha$ be positive operators on $\Hh$  and let
$K^\alpha$ be operators 
such that 
$\sum_\alpha K^{\alpha *} K^{\alpha } \leq \id_\Hh$. Then the map
\begin{equation}\label{la}
(A^1,\dots,A^M) \mapsto \Tr_\Hh \exp\left( L + \sum_{\alpha=1}^M K^{\alpha *} 
(\ln A^\alpha) {K^\alpha}\right)
\end{equation}
is jointly concave.
\end{thm}

\noindent {\bf Remarks.} 
\begin{enumerate}
  
\item This theorem was proved in \cite[Thm.~6]{L} for one $A$ and $\K=\id$, and
  it was generalized there, in Corollary 6.1, to $A^1,\dots,A^M$, but
  only in the case that the $K^\alpha$ are nonnegative numbers
  $\sqrt{p^\alpha}$ with $\sum_\alpha p^\alpha \leq 1$.
  
\item Theorem~\ref{L1} can be extended to the infinite dimensional
  case as well, using the methods of Sect.~4 in \cite{L}. It is also
  possible to generalize to the case of continuous variables $\alpha$
  in some measure space $\Omega$; in this case, $A^\alpha$ is a
  measurable function on $\Omega$ with values in the positive
  operators, and the sum over $\alpha$ in (\ref{la}) is replaced by
  the integral $\int_\Omega d\mu(\alpha) \,K^{\alpha *} (\ln A^\alpha)
  {K^\alpha}$, with $\K$ satisfying $\int_\Omega d\mu(\alpha) \, \KK\K
  \leq \id$. For simplicity we will not give this generalization here
  since we will not need it for the proof of our main
  Theorem~\ref{Tmain}.

\item Note the switching of $\K$ and $\KK$ between
(\ref{rho1a})--(\ref{rho3}) and (\ref{la}). Note also that only the
inequality $\sum_\alpha \KK\K\leq \id$ is required for
Theorem~\ref{L1}, whereas {\it equality} is necessary in
Theorem~\ref{T1}.
\end{enumerate}

\section{Proof of Theorems~\ref{T1} and~\ref{L1}}\label{proofsect}

\begin{proof}[Proof of Theorem \ref{L1}]
  We start with two preliminary remarks. (a.) It is clearly enough to
  assume that $\sum_\alpha K^{\alpha *} K^{\alpha } = \id$, for
  otherwise we can add one more $K^{M+1}=\left(\id - \sum_\alpha
    K^{\alpha *} K^{\alpha } \right)^{1/2}$ and take $A^{M+1}=\id$.
  (b.) We can also assume that all $K^\alpha$ are invertible; the
  general case follows by continuity.

Let $\Kk=\Hh\otimes \C^M$. Every operator in 
$\Bb(\Kk)$ can be thought of as an $M\times M$ matrix 
(indexed by $\alpha, \beta$) with entries in $\Bb(\Hh)$. 
Define $\widehat L$, $\widehat A$, $\widehat P \in \Bb(\Kk)$ by
\begin{equation}
\widehat L_{\alpha\beta} = \frac 1M \delta_{\alpha\beta} 
\frac 1{\KK} L \frac 1{\K},
\end{equation}
\begin{equation}
\widehat A_{\alpha\beta} =  \delta_{\alpha\beta} A^\alpha ,
\end{equation}
and
\begin{equation}
\widehat P_{\alpha\beta} = {\K}\, {K^{\beta *}}.
\end{equation}
Note that $\widehat P = \widehat{P}^* $  and (since 
$\sum_\alpha K^{\alpha *} K^{\alpha } = \id$) 
$\widehat{P}^2 = \widehat P$, so  $\widehat P$  is an 
orthogonal projection. 
We know from \cite[Thm.~6]{L} that the map
\begin{equation}\label{exp}
(A^1,\dots,A^M) \mapsto \Tr_{\Kk} \exp\left( -\lambda (\id-\widehat P) + 
\widehat L + \ln \widehat A\right)
\end{equation}
is concave, for every $\lambda\in \R$. The concavity property survives 
in the limit 
$\lambda\to\infty$, in which limit  the operator  in the exponent
of  (\ref{exp}) is $-\infty$ on the orthogonal complement of $ \widehat P\Kk $.
Therefore, in the $\lambda \to \infty $ limit the concavity in (\ref{exp}) 
becomes the statement that
\begin{equation}
(A^1,\dots,A^M) \mapsto \Tr_{\widehat P\Kk}  \exp\left( \widehat P \left[  
\widehat L + \ln \widehat A\right] \widehat P\right)
\end{equation}
is concave. We shall show that 
\begin{equation}\label{fin}
\Tr_{\widehat P \Kk}  \exp\left( \widehat P \left[  \widehat L + \ln 
\widehat A\right] \widehat P\right) = \Tr_{\Hh} \exp\left(  L + 
\sum_\alpha \KK (\ln A^\alpha) \K  \right),
\end{equation}
which finishes the proof. 

Eq. (\ref{fin}) can be proved as follows.
The trace on the left side is over $\widehat P \Kk$, which is 
isomorphic to $\Hh$. 
In fact, the map $\U : \Hh \to \widehat P \Kk$, defined by
\begin{equation}
\left(\U \Psi \right)^\alpha = \K \Psi
\end{equation}
is clearly onto since every vector in $\widehat P \Kk$ has the form
$\K \Psi$. Moreover, since $\sum_\alpha \KK \K=\id$, $\U$ preserves
norms and hence $\U$ is a unitary.  A simple calculation shows that
\begin{equation}
\U^* \widehat P \left[  \widehat L + \ln \widehat A\right] 
\widehat P \U = L + \sum_\alpha \KK (\ln A^\alpha) \K. 
\end{equation}
\end{proof}

\begin{proof}[Proof of Theorem~\ref{T1} using Theorem~\ref{L1}]
We need to show that 
\begin{equation}\label{eq1}
\Tr_{123}\,  \rho_{123}\left( -\ln \rho_{123} + \ln \rho_{12} - 
\sum_\alpha \KK \left( \ln \rho_2^\alpha- 
\ln \rho_{23}^\alpha \right)\K \right)\leq 0.  
\end{equation}
Using the Peierls-Bogoliubov inequality \cite{Thirring}, we see that (\ref{eq1}) 
holds if we can show that
\begin{equation}\label{eq3}
\Tr_{123} \exp\left( \ln \rho_{12} - \sum_\alpha \KK
\left( \ln \rho_2^\alpha- \ln \rho_{23}^\alpha \right)\K
\right) \leq 1.
\end{equation} 
We now use an idea of  Uhlmann \cite{U}. 
Let $U_3$ be a unitary operator on $\Hh_3$, and let $dU_3$ denote the 
corresponding normalized Haar measure. Since the trace is invariant under 
unitary transformations, and since the $K^\alpha$ commute with $U_3$, we see 
that the left side of (\ref{eq3}) equals
\begin{eqnarray}\nonumber
&&\!\!\!\!\!\!\!\! \!\!\! \int  \Tr_{123}\, U^*_3 \exp\left( \ln \rho_{12} 
-\sum_\alpha \KK \left( \ln \rho_2^\alpha- \ln \rho_{23}^\alpha 
\right)\K \right) U_3 \, dU_3 \ \ = \\ 
\nonumber &&\!\!\!\!\!\!\!\! \!\!\! 
\int  \Tr_{123}  \exp\left( \ln \rho_{12} - \sum_\alpha \KK
\left( \ln \rho_2^\alpha \right)\K + \sum_\alpha \KK 
\left( \ln U_3^* \rho_{23}^\alpha U_3\right) \K \right)dU_3.\\  \label{uu}
\end{eqnarray}
Now $\int  \left[U_3^* \rho_{23}^\alpha U_3 \right]dU_3 = d^{-1} \Tr_3 \, 
\rho_{23}^\alpha 
= d^{-1} \rho_2^\alpha$, where $d$ denotes the dimension of $\Hh_3$. 
Using the concavity result of Theorem~\ref{L1}, we see that 
\begin{equation}
(\ref{uu})\leq \Tr_{123} \, \exp\left( \ln\rho_{12} - \sum_\alpha 
\KK\K (\ln d )\right) = 1.
\end{equation} 
The last equality follows from $\sum_\alpha \KK\K=\id$ and $\Tr_{123}\,  
\rho_{12}=d$. 
\end{proof}

\begin{proof}[Proof of Theorem~\ref{T1} using CPT theory]
Consider the map $\Phi:\Hh_1\otimes \Hh_2 \otimes \Hh_3 
\mapsto \C^M\otimes \Hh_2\otimes\Hh_3$, given for a general density 
matrix $\rho_{123}$ by 
\begin{equation}
\Phi(\rho_{123})=\bigoplus_\alpha n^\alpha \rho_{23}^\alpha.
\end{equation}
This map is trace preserving and completely positive (CPT) \cite{davies}. It is known 
that the relative entropy, $H(\rho,\gamma)=\Tr\, \rho(\ln\rho-\ln\gamma)$ 
decreases under such maps \cite{lindblad,uhlmann,ruskaiJMP}, and hence
\begin{equation}\label{rele}
H(\rho_{123},\rho_{12}\otimes \rho_3) \geq H(\Phi(\rho_{123}), 
\Phi(\rho_{12}\otimes \rho_3)).
\end{equation}
The left side of this inequality equals $S[\rho_{12}]+S[\rho_3]- S[\rho_{123}]$. 
To compute the right side, note that 
\begin{equation}
\Phi(\rho_{12}\otimes\rho_3) = \bigoplus_\alpha n^\alpha \rho_{2}^\alpha \otimes \rho_3.
\end{equation}
It is then easy to see that the right side of (\ref{rele}) 
equals $\sum_\alpha n^\alpha(S[\rho_{2}^\alpha] - S[\rho_{23}^\alpha] 
+ S[\rho_3])$. Thus (\ref{rele}) is the same statement as  (\ref{teq}).   
\end{proof}

\section{Corollaries and Applications}\label{applications}

Taking $\Hh_2=\C$, we get as an immediate corollary of Theorem~\ref{Tmain}
and the concavity of $\rho \mapsto S[\rho]$:

\begin{cor}[Improved Subadditivity]\label{C1}
  Let $\rho_{12}$ be a density matrix on a separable Hilbert space
  $\Hh_1\otimes\Hh_2$. Let $P^\alpha$ be positive, bounded and
  measurable operators on $\Hh_1$, with $\int_\Omega d\mu(\alpha)\, 
  P^\alpha =\id_1$. Let $n^\alpha = \Tr_{12}\, P^\alpha \rho_{12}$
  and, in case $n^\alpha>0$, let $\rho_{2}^\alpha = \Tr_{1}\,
  P^\alpha \rho_{12} / n^\alpha$. Then
\begin{equation}\label{ceq}
S[\rho_{12}] \leq    S[\rho_1] + \int_\Omega d\mu(\alpha) \, n^\alpha  S[\rho_{2}^\alpha]
\leq S[\rho_1] +  S[\rho_2].
\end{equation}
\end{cor}

\noindent {\bf Remarks.} 
\begin{enumerate}
\item In the notation of Theorem~\ref{Tmain}, $P^\alpha= \KK\K$, but
  there is no need for this splitting in this case.
  
\item One may wonder whether (\ref{ceq}) holds if $\rho_1$ is also
  split in a manner similar to $\rho_2$. This is not true, in general!
  As a simple example, consider the case when $\Hh_1=\Hh_2$, and
  $\rho_{12}= d^{-1} \sum_{\alpha=1}^d \Pi^\alpha\otimes \Pi^\alpha$,
  with $\Pi^\alpha$ being mutually orthogonal one-dimensional
  projections. With $P^\alpha = \Pi^\alpha$ we have $S[\rho_{12}] =
  \ln d$, whereas $S[\rho_1^\alpha]=S[\rho_2^\alpha]=0$ for all
  $\alpha$.
\end{enumerate}

\subsection{Classical Entropies}\label{sect31}

Now, suppose we are given a partition of unity of both $\Hh_1$ and
$\Hh_2$, i.e., a finite set of positive operators $P^\alpha$ and
$Q^\beta$ such that
\begin{equation}\label{deco}
\sum_\alpha P^{\alpha} = \id_1 \ , 
\quad \sum_\beta Q^{\beta} = \id_2.
\end{equation}
For simplicity, we restrict ourselves to the case of a finite
dimensional Hilbert space and discrete sums in this subsection, but,
using the methods described in the appendix, one can extend the
results to the case of infinite dimensional (separable) Hilbert spaces
and integrals over general measure spaces.

For $\rho_{12}$ a density matrix on $\Hh_1\otimes \Hh_2$, we can define a 
\lq classical\rq\ entropy as 
\begin{equation}
S^{\rm cl}[\rho_{12}] = \sum_{\alpha,\beta} - \big( \Tr_{12}\, P^\alpha 
Q^\beta\rho_{12}  \big) \ln  \big( \Tr_{12}\, P^\alpha 
Q^\beta\rho_{12} \big).
\end{equation}
Analogously, we can define $S^{\rm cl}[\rho_1]$ and $S^{\rm cl}[\rho_2]$ 
for density matrices on $\Hh_1$ or $\Hh_2$. 

In the case where the $P^\alpha$ and $Q^\beta$ are one-dimensional
projections, this definition agrees with the one in \cite{wehrlA}.
Note, however, that we define the classical entropy here for any
partition of unity. If this partition is trivial, i.e., $P^\alpha=\id$
for $\alpha=1$ and zero otherwise, then $S^{\rm cl}[\rho_1]\equiv 0$
identically.  In particular, it is in general not true that
$S[\rho_{1}]\leq S^{\rm cl} [\rho_{1}]$.  However, this is true if the
partition is such that $\Tr\, P^\alpha\leq 1$ for all $\alpha$, in
which case this inequality follows from concavity of $x\mapsto -x\ln
x$.

Corollary \ref{C1} above can be used to prove the following inequality for 
the mutual information $S[\rho_1]+S[\rho_2]-S[\rho_{12}]$.

\begin{cor}[Quantum mutual information bounds classical mutual information]\label{C2}
  For any density matrix $\rho_{12}$ on $\Hh_1\otimes\Hh_2$, and any
  partition of unity as in (\ref{deco}),
\begin{equation}\label{weh}
S[\rho_1]+S[\rho_2]-S[\rho_{12}] \geq    
S^{\rm cl}[ \rho_{1}] +  S^{\rm cl}[ \rho_{2}] - S^{\rm cl}[\rho_{12}] . 
\end{equation}
\end{cor}

Note that the right side of (\ref{weh}) is just the classical mutual
information, since $\Tr_1\, P^\alpha \rho_1 = \sum_\beta
\Tr_{12}\, P^\alpha Q^\beta \rho_{12}$.

\begin{proof}
We learn from Corollary~\ref{C1} that
\begin{equation}\label{ls}
S[\rho_{12}]-S[\rho_1]  \leq \sum_\alpha n^\alpha S[\rho_2^\alpha] ,
\end{equation}
where $n^\alpha=\Tr_{12}\, P^\alpha \rho_{12}$ and 
$\rho_2^\alpha = \Tr_1\, P^\alpha \rho_{12} / n^\alpha$. 
On $\C^M\otimes \Hh_2$, define a density matrix $\widetilde \rho_{12}$ as
\begin{equation}
\widetilde \rho_{12}= \bigoplus_\alpha n^\alpha \rho_2^\alpha.
\end{equation} 
Then the right side of (\ref{ls}) can be written as
\begin{equation}
\sum_\alpha n^\alpha S[\rho_2^\alpha] = S[\widetilde\rho_{12}] -  S^{\rm cl}[\rho_1].
\end{equation}
Note that $\widetilde\rho_2= \Tr_1\, \widetilde \rho_{12}
= \sum_\alpha n^\alpha \rho_2^\alpha=\rho_2$.

We now apply inequality Corollary~\ref{C1} again, this time to the expression 
$S[\widetilde\rho_{12}]-S[\widetilde\rho_2]$. This yields 
\begin{equation}
S[\widetilde \rho_{12}]-S[\rho_2] = S[\widetilde\rho_{12}] - S[\widetilde\rho_2] 
\leq \sum_\beta m^\beta S[\widetilde \rho_1^\beta],
\end{equation}
with $m^\beta= \Tr_{12}\, Q^\beta \widetilde \rho_{12} $ and 
$\widetilde \rho_1^\beta = \Tr_2\, Q^\beta\widetilde\rho_{12}/m^\beta$. 
The right side of this expression equals
\begin{equation}\label{ls2}
\sum_\beta m^\beta S[\widetilde \rho_1^\beta] = S^{\rm cl}[\rho_{12}]-S^{\rm cl}[\rho_2].
\end{equation}
In combination (\ref{ls})--(\ref{ls2}) give the desired result.
\end{proof}

Another way to interpret the results above is the following: define a partially 
classical and partially quantum entropy by
\begin{equation}
S^{\rm cl, Q}[\rho_{12}]= - \sum_\alpha \Tr_2\, \big(\Tr_1\, P^\alpha \rho_{12}\big) 
\ln  \big(\Tr_1\, P^\alpha \rho_{12}\big).
\end{equation}
Then Corollary~\ref{C1} and the proof of Corollary~\ref{C2} show that
\begin{eqnarray}\nonumber
S[\rho_{12}]-S[\rho_1]-S[\rho_2] &\leq& S^{\rm cl,Q}[\rho_{12}]-S^{\rm cl}[\rho_{1}]
-S[\rho_{2}]\\&\leq& S^{\rm cl}[\rho_{12}]-S^{\rm cl}[\rho_1]-S^{\rm cl}[\rho_2].
\end{eqnarray}
In the same way, Theorem~\ref{T1}, in the special case where the $\K$
act non-trivially only on $\Hh_1$, can be interpreted as
\begin{equation}\label{cqq}
S[\rho_{123}]-S[\rho_{12}]\leq S^{\rm cl,Q,Q}[\rho_{123}]-S^{\rm cl,Q}[\rho_{12}]
\end{equation} 
for a partition of unity on $\Hh_1$, with the obvious definition of
$S^{\rm cl,Q,Q}$.  We can use this inequality to prove the following.

\begin{cor}[Convexity of classical minus quantum entropy]\label{C3}
The map
\begin{equation}
\rho_{12} \mapsto S^{\rm cl,Q}[\rho_{12}]- S[\rho_{12}] 
\end{equation}
is convex.
\end{cor}

\begin{proof}
Let $A_{12}$ and $B_{12}$ be two density matrices on $\Hh_1\otimes \Hh_2$. 
On $\Hh_1\otimes\Hh_2\otimes \C^2$, consider the density matrix 
$\rho_{123} = \half A_{12}\otimes \Pi + \half B_{12}\otimes (\id-\Pi)$, 
where $\Pi$ is a one-dimensional projection in $\C^2$. Inequality (\ref{cqq}) 
implies that 
\begin{equation}
\half S[A_{12}]+\half S[B_{12}] - S[\rho_{12}] \leq \half S^{\rm cl,Q}[A_{12}]
+\half S^{\rm cl,Q}[B_{12}] - S^{\rm cl,Q}[\rho_{12}]. 
\end{equation}
Since $\rho_{12}=\half A_{12}+\half B_{12}$, this proves the result.
\end{proof}

{\noindent {\bf Remark.} In particular, taking $\Hh_2=\C$ to be
  trivial, this corollary shows that the map (for a single Hilbert
  space)
\begin{equation}\label{convex}
\rho\mapsto S^{\rm cl}[\rho]-S[\rho]
\end{equation}
is convex -- which is remarkable, given that both entropies are
concave functions of $\rho$. The inequality implied by convexity of
(\ref{convex}) is known as the Holevo bound \cite{hol1,hol2}.

\subsection{Coherent States and Wehrl Entropy}\label{cohsect}

Now, suppose we are given a coherent state decomposition of both
$\Hh_1$ and $\Hh_2$, i.e., normalized vectors
$|\varphi\rangle\in\Hh_1$, $|\theta\rangle\in\Hh_2$ and positive
measures $\mu$ and $\nu$ on some measure space (not necessarily the
same spaces) such that 
\begin{equation}\label{deco2} \int d\mu(\varphi)\, 
|\varphi\rangle\langle\varphi| = \id_1 \ , \ \int d\nu(\theta)\,
|\theta\rangle\langle\theta| = \id_2.  
\end{equation} 
Here, $|\varphi\rangle\langle\varphi|$ is the Dirac notation for the
one-dimensional projector onto $\varphi$ and the integrals are to be
interpreted in the weak sense, as explained before.
  
The classical (Wehrl) entropy \cite{wehrl} for a density matrix $\rho_{12}$ 
on $\Hh_1\otimes
\Hh_2$ is then defined as 
\begin{equation} 
S^{\rm W}[\rho_{12}] = - \int d\mu(\varphi) d\nu(\theta)\, \langle
\varphi,\theta|\rho_{12}|\varphi, \theta\rangle \ln \langle
\varphi,\theta|\rho_{12}|\varphi,\theta\rangle,
\end{equation}
and similarly for density matrices on $\Hh_1$ or
$\Hh_2$. As Wehrl showed, it follows from concavity of $x\mapsto
-x\ln x$ that
\begin{equation}\label{wehrl2} 
S[\rho_{12}]\leq S^{\rm W} [\rho_{12}].
\end{equation} 

Corollaries \ref{C2} and \ref{C3} now also hold for the Wehrl entropy.
In particular, it is true that
\begin{equation}
 S^{\rm W}[\rho_1] + S^{\rm W}[\rho_2] - S^{\rm W}[\rho_{12}]
\leq  S[\rho_1] + S[\rho_2] - S[\rho_{12}] .  
\end{equation}
Moreover, Corollary~\ref{C3} implies that the map 
\begin{equation}
\rho\mapsto S^{\rm W}[\rho]-S[\rho]
\end{equation}
is convex.  Note that the infimum of this function is zero. In the
finite dimensional case, this infimum is achieved for the totally
mixed state $\rho= d^{-1} \id$, where $d={\rm dim\, } \Hh$.

It might be recalled that Wehrl raised the question \cite{wehrl} of
evaluating the minimum of his classical entropy and conjectured, in
the special case of the Glauber coherent states, that it should be
given by the one-dimensional projector onto a coherent state $\rho =
|\theta \rangle \langle \theta|$. (The minimum of the quantum entropy,
$-\Tr \ \rho \ln \rho$, is always zero.)  This particular conjecture
was proved in \cite{liebcoh}, where the (still open) generalized
conjecture was made to $SU(2)$ (Bloch) coherent states.  Oddly, the minimum of
the \textit{difference} of the entropies is a much easier question to
answer.

Note that because of convexity the maximum of the function $S^{\rm
  W}[\rho] - S[\rho]$ is attained for a pure state, where $S[\rho]=0$.
Hence the question about the maximum of $S^{\rm W}[\rho] - S[\rho]$ is
equivalent to {\it maximizing} $S^{\rm W}[\rho]$ over pure states. In
the case of the Glauber coherent states, this maximum is infinite
\cite{wehrl}.

\subsection{Quantum Statistical Mechanics of Point Particles}\label{Cfsect}

Consider a system of two types of particles, $A$ and $B$. The state
space of the combined system is $\Hh=\Hh_A\otimes \Hh_B$, where
$\Hh_A$ and $\Hh_B$ are the spaces of square integrable functions of
the particle configurations of particles $A$ and $B$, respectively. We
assume that the configuration space is $\R^{d_A}$ and $\R^{d_B}$,
respectively, for some finite $d_A$ and $d_B$.  The usual subadditivity of
entropy implies that, for any state $\rho$ on $\Hh$,
\begin{equation}\label{subb}
S[\rho_B] \geq S[\rho] - S[\rho_A],
\end{equation}
where $\rho_A$ and $\rho_B$ denote the states of the subsystems.
However, in applications it can be useful to get a lower bound not
only on the entropy of $\rho_B$, which is the state averaged over all
configurations of the $A$ particles, but rather on the {\it average
  entropy} of the state for {\it fixed} $A$ particles. Such a bound is
one of the key ingredients in a rigorous upper bound on the pressure
of a dilute Fermi gas at non-zero temperature \cite{S04}.

More precisely, if $X_A$ and $X_B$ denote particle configurations of
the $A$ and $B$ particles, any density matrix on $\Hh$ will be given
by an integral kernel $\rho(X_A,X_B;X_A',X_B')$. For every fixed
configuration of the $A$ particles, $X_A$, we can then define a
density matrix on $\Hh_B$ by the kernel
\begin{equation}\label{defrb}
\rho_B^{X_A}(X_B,X_B') =  n(X_A)^{-1} \, \rho(X_A,X_B;X_A,X_B'),
\end{equation}
where $n(X_A)$ is the normalization factor
\begin{equation}
n(X_A)= \int d X_B \,  \rho(X_A,X_B;X_A,X_B).
\end{equation}
Since $\rho$ is a trace class operator, $\rho_B^{X_A}$ is well defined
by the spectral decomposition of $\rho$ for almost every $X_A$, if
$n(X_A)\neq 0$.  The definition~(\ref{defrb}) makes sense only if
$n(X_A)>0$; only in this case $\rho_B^{X_A}$ will be needed below,
however.

Note that $n(X_A)$ is the probability density of a configuration of
$A$ particles $X_A$.  Moreover, $\int dX_A\, n(X_A) = 1$, and $\int
dX_A\, n(X_A)\, \rho_B^{X_A} = \rho_B$. Therefore we have, by
concavity of $S[\rho]$,
\begin{equation}
S[\rho_B] \geq \int dX_A\, n(X_A)\, S[\rho_B^{X_A}].
\end{equation}
Hence the following is a strengthening of (\ref{subb}). 

\begin{cor}[Subadditivity with Average Entropy instead of Entropy of Average]\label{Cf}
  Let $\rho$ be a density matrix on $\Hh_A\otimes \Hh_B$ with finite
  entropy. With the definitions given above,
\begin{equation}\label{Cfeq}
\int dX_A\, n(X_A)\, S[\rho_B^{X_A}] \geq  S[\rho] - S[\rho_A].
\end{equation}
\end{cor}

We remark that it is not necessary to have a fixed particle number for
this bound; the integral $\int dX_A$ can as well include a discrete
sum over different particle numbers. I.e., our bound also applies to
the grand-canonical ensemble, and this is the form that Corollary~\ref{Cf}
actually gets used in \cite{S04}. For simplicity, we consider only the
case of a fixed particle number in the proof below, but the extension
is straightforward, using an additional decomposition $\id =
\sum_{n\geq 0} P_n$, where $P_n$ projects onto the subspace of $\Hh_A$
with fixed particle number $n$.

Corollary~\ref{Cf} follows from Corollary~\ref{C1} by the following
limiting argument.

\begin{proof}[Proof of Corollary~\ref{Cf}]
  For $d\equiv d_A$, let $j : \R^{d}\mapsto \R$ be a positive and
  integrable function on the configuration space of the $A$ particles,
  with $\int dX\, j(X)=1$. For some $\eps>0$ and $Y\in\R^{d}$, let
  $j_\eps^Y(X)=\eps^{-d}j((X-Y)/\eps)$. Let
\begin{equation}
P^\eps = \id - \int_{\R^{d}} dY\, |j_\eps^Y\rangle\langle j_\eps^Y|.
\end{equation}
It is not difficult to see that $P^\eps\geq 0$. Hence we can infer
from Corollary~\ref{C1} that, for any density matrix $\rho_{12}$ on
$\Hh_1\otimes \Hh_2$ (where we use again the notation $1$ and $2$
instead of $A$ and $B$),
\begin{equation}\label{epsn}
S[\rho_{12}] - S[\rho_1] \leq \int_{\R^{d}} dY\, n_\eps(Y) S[\rho_\eps^Y] 
+ n_{P^\eps} S[\rho_2^{P^\eps}],
\end{equation}
where we denoted $n_\eps(Y) = \langle
j_\eps^Y|\rho_1|j_\eps^Y\rangle$, $\rho_\eps^Y = \langle
j_\eps^Y|\rho_{12}|j_\eps^Y\rangle / n_\eps(Y)$, $n_{P^\eps}= \Tr_{1}
P^\eps \rho_1$ and $\rho_2^{P^\eps} = \Tr_{1} P^\eps \rho_{12} /
n_{P^\eps}$. We will show that there exists a sequence $\eps_j$ with
$\eps_j\to 0$ as $j\to\infty$ such that the right side of (\ref{epsn})
converges to the left side of (\ref{Cfeq}) in the limit $j\to \infty$.

We note that, for any square integrable function $\phi$ on $\R^d$,
$\langle j_\eps^Y|\phi\rangle \to \phi(Y)$ strongly in $L^2(\R^d)$ as
$\eps \to 0$ \cite[Thm.~2.16]{anal}. Passing to a subsequence, it is
then true that $\langle j_\eps^Y|\phi\rangle \to \phi(Y)$ almost
everywhere. Decomposing $\rho_1$ into its eigenvalues and
eigenfunctions, we see that there is a subsequence such that
$\lim_{\eps\to 0} n_\eps(Y) = n(Y)$ for almost every $Y$. Also,
$\rho_\eps^Y \rightharpoonup\rho_B^{Y}$ weakly as $\eps\to 0$ for a.e.
$Y$. (Here we used the separability of the Hilbert space to ensure the
existence of this subsequence.)  Since also the traces converge, this
convergence is actually in trace class norm \cite{wehrl3}.

We first assume that $\rho_{12}$ has finite rank. Then also
$\rho_\eps^Y$ and $\rho_B^Y$ have finite rank, and hence bounded
entropy. It follows that $\lim_{\eps \to 0} S[ \rho_\eps^Y] =
S[\rho_B^Y]$. Moreover, it is easy to see that $P^\eps \rightharpoonup
0$ weakly as $\eps\to 0$.  This implies that $\lim_{\eps\to 0}
n_{P^\eps}=0$, and hence $\lim_{\eps\to 0} \int dY\, n_\eps(Y) =1$.
It then follows from Fatou's Lemma that
\begin{equation}
\lim_{\eps\to 0} \int_{\R^d} dY\, n_\eps(Y) S[\rho_\eps^Y] 
= \int_{\R^d} dY\, n(Y) S[\rho_B^Y].
\end{equation}

It remains to show that the last term in (\ref{epsn}) goes to zero as
$\eps\to 0$. As already noted, $\lim_{\eps\to 0} n_{P^\eps}=0$.  The
entropy $S[\rho_2^{P^\eps}]$ need not be bounded as $\eps\to 0$,
however. Since $P^\eps\leq \id$, $\Tr_{1} P^\eps \rho_{12}\leq
\rho_B$.  Note that $\rho_{12}$ has finite entropy by assumption and,
without loss of generality, also $\rho_A$ has finite entropy. This
implies that $\rho_B$ has finite entropy by the triangle inequality
for entropies \cite{LAraki}. Hence it follows from dominated convergence
\cite[Thm.~A3]{LR} that $S[ \Tr_{1} P^\eps \rho_{12}]\to 0$ as $\eps\to
0$, and hence $n_{P^\eps} S[\rho_2^{P^\eps}] = S[\Tr_{1} P^\eps
\rho_{12}] + n_{P^\eps} \ln n_{P^\eps} \to 0$ as $\eps \to 0$. This
proves (\ref{Cfeq}) in the case that $\rho_{12}$ has finite rank.

For a general $\rho_{12}$ on $\Hh_1\otimes \Hh_2$, let
$\rho^j_{12}=P^j\rho_{12}$, where $P^j$ denotes the projection onto
the largest $j$ eigenvalues of $\rho_{12}$. It is then easy to see that
\begin{equation}
S[\rho_{12}]= \lim_{j\to\infty} S[\rho_{12}^j]
\end{equation}
and
\begin{equation}
S[\rho_{A}]= \lim_{j\to\infty} S[\Tr_2\, \rho_{12}^j]
\end{equation}
(cf. the Appendix in \cite{LR}). Moreover, with $n^j(Y)$ and
$\rho_{B,j}^Y$ defined as above, for the operator $\rho_{12}^j$, we write
\begin{eqnarray}\nonumber
\int dY \, n^j(Y) S[\rho_{B,j}^Y] &=& e \int dY\, n(Y) S\big[\rho_{B,j}^Y 
\big(n^j(Y)/e\, n(Y)\big)\big] \\ && - \int dY\, n^j(Y) 
\ln \big( e\, n(Y) / n^j(Y)\big).
\label{rsft}
\end{eqnarray}
Note that $n^j(Y)$ is pointwise increasing in $j$, and also $n^j(Y)
\rho_{B,j}^Y$ is an increasing sequence of operators. Moreover,
$\rho_{B,j}^Y(n^j(Y)/e\, n(Y))\leq 1/e$. Since $-x\ln x$ is monotone
increasing for $0\leq x\leq 1/e$, this implies that the first term on
the right side of (\ref{rsft}) is bounded from above by
\begin{equation}
e \int dY\, n(Y) S\big[\rho_B^Y /e\big]= \int dY\, n(Y) S\big[\rho_B^Y\big] + \int dY\, n(Y).
\end{equation}
Moreover, since $\lim_{j\to\infty} n^j(Y) = n(Y)$ for almost every
$Y$, this implies, by monotone convergence,
\begin{equation}
\lim_{j\to\infty} - \int dY\, n^j(Y) \ln \big( e\, n(Y) / n^j(Y)\big) =  
- \int dY\, n(Y) .
\end{equation}
This shows that (\ref{Cfeq}) also holds in the infinite rank case, and
finishes the proof of Corollary~\ref{Cf}.
\end{proof}

\appendix
\section{Extension to Integrals}\label{intapp}

In this appendix we extend Theorem~\ref{T1} in the following way. Let
again $\Hh_i$ be {\it finite dimensional} Hilbert spaces, $1\leq i\leq 3$.
Let $\Omega$ be a measure space, with elements labeled by $\alpha$,
and let $d\mu(\alpha)$ be a measure on $\Omega$. Let $K^\alpha$ be
matrices on $\Hh_1\otimes \Hh_2$ that are weakly measurable 
(i.e., all the matrix elements are
measurable functions), such that 
\begin{equation}\label{inteq} 
\int_\Omega
d\mu(\alpha\,) \KK \K = \id_{12}.  
\end{equation} 
The extension of Theorem~\ref{T1} is the following. With the same 
definitions as in (\ref{rho1a})--(\ref{rho3}),
\begin{equation}\label{teq2}
S[\rho_{123}] - S[\rho_{12}] \leq    
\int_\Omega d\mu(\alpha) \,  n^\alpha \big(  S[\rho_{23}^\alpha]
- S[\rho_2^\alpha] \big).
\end{equation}
Note that this expression is well defined, since the integrand is
measurable and the entropy is bounded.

For the proof of (\ref{teq2}), we may assume that, for each $\alpha$,
$\|\KK\K\|\leq 1$. This is no restriction, since we can always absorb
the norm into the measure $d\mu(\alpha)$. Likewise, we may assume that
$\Tr\, \KK\K \geq 1/2$. Taking the trace of (\ref{inteq}), it is then
clear that $\Omega$ has finite measure.

Pick some $\eps>0$. By looking at the level sets of the matrix
elements of $K^\alpha$, we can divide $\Omega$ into {\it finitely
  many} disjoint measurable subsets $\Oo_j$, $1\leq j\leq M_\eps$,
with $\| \K - K^\beta\| \leq \eps$ if $\alpha$ and $\beta$ are in the
same subset $\Oo_j$. For each $\alpha\in \Omega$, write $K^\alpha =
U^\alpha (\KK\K)^{1/2}$, with $U^\alpha$ unitary. For each $j$, pick
some $\alpha_j \in \Oo_j$, and define
\begin{equation}
L^j = U^{\alpha_j} \left( \int_{\Oo_j} d\mu(\alpha)\, \KK\K \right)^{1/2}.
\end{equation}
We then have $\sum_j L^{j*} L^j = \id_{12}$, and hence we can apply
Theorem~\ref{T1}. That is, we have, for any density matrix
$\rho_{123}$ on $\Hh_1\otimes \Hh_2\otimes \Hh_3$,
\begin{equation}\label{rsp}
S[\rho_{123}] - S[\rho_{12}] \leq \sum_j n^j \left (S[\rho_{23}^j] 
- S[\rho_2^j]\right),
\end{equation}
with $n^j = \Tr_{123} L^{j*} L^j \rho_{123}$ and $\rho_{23}^j =
\Tr_{1} L^j \rho_{123} L^{j*} /n^j$. We will now show that, as
$\eps\to 0$, the right side of (\ref{rsp}) converges to the right side
of (\ref{teq2}).

Using the fact that $\|\sqrt A - \sqrt B\|\leq \|A-B\|^{1/2}$ for any
two positive matrices $A$ and $B$ \cite[Eq.~X.2]{Bhatia}, we can
estimate
\begin{equation}
\| L^j - |\Oo_j|^{1/2} K^{\alpha_j}\|^2 \leq  \int_{\Oo_j} d\mu(\alpha) \| 
\KK\K - K^{\alpha_j*} K^{\alpha_j} \| \leq 2 \eps |\Oo_j|. 
\end{equation}
Here, $|\, \cdot \, |$ denotes the measure of a subset of $\Omega$,
and we used $\|\K\|\leq 1$ and $\|\K-K^{\alpha_j}\|\leq \eps$ in the
last step. Using the triangle inequality we thus see that $\| L^j -
|\Oo_j|^{1/2} K^{\alpha}\|\leq |\Oo_j|^{1/2}(\eps + \sqrt {2\eps})$
for any $\alpha\in \Oo_j$. This implies that
\begin{equation}\label{reff}
\|  L^{j} \rho_{123}L^{j*}  - |\Oo_j|\, K^{\alpha} \rho_{123} K^{\alpha *} 
\| \leq  |\Oo_j| 2(\eps + \sqrt{2\eps}) \|\rho_{123}\|
\end{equation}  
for any $\alpha\in \Oo_j$, where we again used that $\|\K\|\leq 1$.
Note that (\ref{reff}) implies the same estimate for the partial trace
of the operator on the left side, with the right side multiplied by
the dimension of the space. Moreover, since in finite dimensions the
entropy is Lipschitz continuous, this also implies that, for some
constant $C>0$,
\begin{eqnarray}\nonumber
&&\!\!\!\!\!\!\!\sum_j n^j \left (S[\rho_{23}^j] - S[\rho_2^j]\right) 
= \sum_j  \left (S[\Tr_1\, L^{j} \rho_{123}L^{j*}] - S[\Tr_{12}\, L^{j} 
\rho_{123}L^{j*}]\right) 
\\ \nonumber &&\!\!\!\!\!\!\! \leq \sum_j \int_{\Oo_j} d\mu(\alpha)  
\left (S[\Tr_1\, \K\rho_{123} \KK] - S[\Tr_{12}\, \K \rho_{123}\KK] 
+ C(\eps + \sqrt{2\eps}) \right) \\ &&\!\!\!\!\!\!\! = \int_\Omega d\mu(\alpha) \,  
n^\alpha \big(  S[\rho_{23}^\alpha]- S[\rho_2^\alpha] \big) 
+ C |\Omega| (\eps + \sqrt{2\eps}).
\end{eqnarray}
Letting $\eps\to 0$ this proves the desired result.

\section{Extension to Infinite Dimensional Spaces}\label{infapp}

We now show that (\ref{teq2}) holds in the case when the $\Hh_i$ are
separable, possibly infinite dimensional, Hilbert spaces. This will
prove Theorem~\ref{Tmain}. As stated there, we assume that the
$K^\alpha$ are bounded, weakly measurable operators, i.e, $\langle
\phi | K^\alpha|\phi\rangle$ is measurable for all vectors
$|\phi\rangle$.  (By polarization, this implies that the matrix
elements $\langle \phi | K^\alpha|\psi\rangle$ are measurable for any
vectors $|\phi\rangle$ and $|\psi\rangle$.)  We also assume
(\ref{enmain}) to hold. Again it is then no restriction to assume that
$\|\KK\K\|\leq 1$ for each $\alpha \in \Omega$. Note that $\Omega$
need not have finite measure in this case, however. We assume that the
density matrix $\rho_{123}$ has finite entropy. We may also assume
that $S[\rho_{12}]$ is finite, otherwise there is nothing to prove.
Note that in this case, the triangle inequality for entropy
\cite{LAraki} implies that also $S[\rho_3]$ is finite.

For $m\in\N$, let $P_i^{(m)}$ be rank $m$ projections in $\Hh_i$, such
that $P_i^{(m)}\to \id_{\Hh_i}$ strongly as $m\to\infty$. Let
$\widehat P=P_1^{(m)}\otimes P_2^{(m)}\otimes P_3^{(m)}$. Then also
$\widehat P\to \id_{123}$ strongly as $m\to \infty$. In the following,
we find it convenient to suppress the dependence on $m$ in our
notation, but rather put a $\widehat{\phantom{x}}$ on all quantities
that depend on $m$.

Given a density matrix $\rho_{123}$, we define $\widehat \rho_{123} =
\widehat P \rho_{123} \widehat P$. Moreover, let $\widehat \K =
P_1^{(m)}\otimes P_2^{(m)} \K P_1^{(m)}\otimes P_2^{(m)}$ for
$\alpha\in\Omega$.  Let also
\begin{equation}
\widehat L = \left[ P_1^{(m)}\otimes P_2^{(m)} \left( 1 - \int_\Omega d\mu(\alpha) \, 
\KK P_1^{(m)}\otimes P_2^{(m)} \K \right)  P_1^{(m)}\otimes P_2^{(m)} \right]^{1/2} .
\end{equation}
If we set $\widehat \Hh_i = P_i^{(m)} \Hh_i$, we then have
\begin{equation}
\int_\Omega d\mu(\alpha)\, \widehat \KK \widehat \K + \widehat L^2 = \id_{\widehat \Hh_{12}},
\end{equation}
and hence we can apply the finite dimensional result (\ref{teq2}). I.e.,
\begin{equation}\label{tre3}
S[\widehat \rho_{123}] - S[\widehat \rho_{12}] \leq    
\int_\Omega d\mu(\alpha) \,  \widehat n^\alpha \big(  S[\widehat \rho_{23}^\alpha]
- S[\widehat \rho_2^\alpha] \big) +  
\widehat n^L \big( S[\widehat \rho_{23}^L]- S[\widehat \rho_2^L]\big).
\end{equation}
Here, $\widehat n^\alpha$ and the density matrices $\widehat
\rho_{23}^\alpha$ and $\widehat \rho_{2}^\alpha$ are defined as in
(\ref{rho1a})--(\ref{rho3}), with $\K$ replaced by $\widehat\K$ and
$\rho_{123}$ replaced by $\widehat \rho_{123}$. Moreover, $\widehat
\rho_{23}^L$, $\widehat
\rho_{2}^L$ and $\widehat n^L$ are defined in the same way, with
$\widehat L$ in place of $\widehat \K$. Our goal is to show that we
can remove the $\widehat{\phantom{x}}$'s in (\ref{tre3}).

Note that $\widehat \rho_{123}\to \rho_{123}$ strongly as
$m\to\infty$. Since also the trace of $\widehat \rho_{123}$ converges
to $\Tr \rho_{123} =1$, this implies that the convergence is actually
in trace norm, as proved by Wehrl in~\cite{wehrl3}. Hence also
$\widehat \rho_{12} \to \rho_{12}$ in trace norm. Since the
eigenvalues of $\widehat \rho_{123}$ are smaller than the
corresponding eigenvalues of $\rho_{123}$, i.e., $\widehat \rho_{123}
\nl \rho_{123}$ in Simon's notation in the appendix of \cite{LR}, Theorem~A2 in \cite{LR}
implies that
\begin{equation}
\lim_{m\to \infty} S[\widehat\rho_{123}] = S[\rho_{123}].
\end{equation}
By the same reasoning, this also holds for $\widehat\rho_{12}$. 
Taking the limit $m\to\infty$ in (\ref{tre3}), we thus have
\begin{eqnarray}\nonumber 
S[\rho_{123}] - S[\rho_{12}] &\leq&   \liminf_{m\to \infty}  
\int_\Omega d\mu(\alpha) \left( A_m(\alpha) + B_m(\alpha) \right) \\ 
&& + \liminf_{m\to \infty}\,  \widehat n^L \big( S[\widehat \rho_{23}^L]
- S[\widehat \rho_2^L]\big) . \label{tre4}
\end{eqnarray}
Here we defined the functions $A_m(\alpha)$ and $B_m(\alpha)$ by
\begin{equation}\label{defan}
A_m(\alpha) =  \widehat n^\alpha \big(  S[\widehat \rho_{23}^\alpha]
- S[\widehat \rho_2^\alpha] - S[\widehat \rho_3^\alpha]\big)
\end{equation}
and
\begin{equation}\label{defbn}
B_m(\alpha) =  \widehat n^\alpha S[\widehat \rho_3^\alpha],
\end{equation}
with $\widehat \rho_3^\alpha= \Tr_2\, \widehat \rho_{23}^\alpha=
\Tr_{12}\, \widehat \KK\widehat\K \widehat\rho_{123}/\widehat
n^\alpha$. The reason for the splitting into the two parts
(\ref{defan}) and (\ref{defbn}) is that $A_m(\alpha)$ is negative,
which allows for the use of Fatou's Lemma, whereas $B_m(\alpha)$ depends
on $\widehat\K$ only through $\widehat\KK\widehat\K$.

We start by estimating the last term on the right side of
(\ref{tre4}). By subadditivity of entropy, $S[\widehat
\rho_{23}^L]- S[\widehat \rho_2^L] \leq S[\widehat \rho_3^L]$, with
$\widehat \rho_3^L = \Tr_2 \widehat \rho_{23}^L$. We claim that
$\lim_{m\to\infty} \widehat n^L = 0$. This is true if we can show that
\begin{equation}\label{ll}
\lim_{m\to\infty} \int_\Omega d\mu(\alpha) \, \Tr_{123} \widehat P \KK 
\widehat P \K \widehat P \rho_{123}  = 1.
\end{equation}
Since $\widehat P \KK \widehat P \K \widehat P$ converges strongly to
$\KK\K$ for each fixed $\alpha$ (because products of strongly
convergent sequences converge strongly), we see that $\Tr_{123}
\widehat P \KK \widehat P \K \widehat P \rho_{123}$ converges to
$\Tr_{123} \KK \K \rho_{123}$ for each fixed $\alpha$. Hence, using Fatou's
Lemma, we see that the left side of (\ref{ll}) is always $\geq
1$. On the other hand, estimating the middle $\widehat P$ in the
integrand in (\ref{ll}) by $\widehat P \leq 1$ and using
(\ref{enmain}), we see that the left side of (\ref{ll}) is bounded
above by
\begin{equation}
\lim_{m\to\infty} \Tr_{123}\, \widehat P \rho_{123} = \Tr_{123}\, \rho_{123} = 1.   
\end{equation}
This proves the claim that $\lim_{m\to\infty} \widehat n^L=0$. 

Although $S[\widehat \rho_3^L]$ need not be bounded, we claim that
$\widehat n^L S[\widehat \rho_3^L]\to 0$ as $m\to\infty$. To see this,
write $\widehat n^L S[\widehat \rho_3^L] = S[\Tr_{12} \widehat L^2
\widehat \rho_{123}] + \widehat n^L \ln \widehat n^L$. Note that the
second term goes to zero as $\widehat n^L\to 0$.  Since $\widehat L^2
\leq \id_{12}$, $\Tr_{12} \widehat L^2 \widehat \rho_{123} \leq P^{(m)}_3\rho_3 P^{(m)}_3 \nl
\rho_3$. Recall that $S[\rho_3]$ is finite. Since $\Tr_{12} \widehat
L^2 \widehat \rho_{123} \to 0$ in trace norm as $m\to \infty$, we can
use dominated convergence \cite[Thm.~A1]{LR} to conclude that
$S[\Tr_{12} L^2 \widehat \rho_{123}]\to 0$ as $m\to \infty$.  Hence we
have shown that
\begin{equation}
\limsup_{m\to \infty} \, \widehat n^L \big( S[\widehat \rho_{23}^L]
- S[\widehat \rho_2^L]\big) \leq 0. 
\end{equation}

Next we treat the first term on the right side of (\ref{tre4}), i.e.,
the integral of $A_m(\alpha)$.  Since $A_m(\alpha)\leq 0$ (by
subadditivity), we can use Fatou's Lemma to estimate 
\begin{equation}\label{65}
\liminf_{m\to\infty} \int_\Omega d\mu(\alpha)\, A_m(\alpha) \leq
\int_\Omega d\mu(\alpha) \limsup_{m\to\infty} A_m(\alpha) .  
\end{equation} 
We now claim that $\widehat \rho_{23}^\alpha \to \rho_{23}^\alpha$ in
trace norm.  It is clear that $\widehat \KK \widehat \rho_{123}
\widehat \K$ converges strongly to $\KK\rho_{123}\K$.  By the same
argument as that after (\ref{ll}), the trace also converges, and hence
the convergence is in trace norm. This implies that the reduced
density matrices also converge in trace norm, and hence proves our
claim.  Moreover, we claim that, for each fixed $\alpha$,
\begin{equation} \label{66}
\limsup_{m\to\infty}\, A_m(\alpha) \leq
n^\alpha \big( S[\rho_{23}^\alpha]- S[\rho_2^\alpha] -
S[\rho_3^\alpha]\big).  
\end{equation} 
This follows from upper semicontinuity of
$S[\rho_{23}]-S[\rho_2]-S[\rho_3]$ in $\rho_{23}$. This upper
semicontinuity, in turn, follows from lower semicontinuity of the
relative entropy $H(\rho,\sigma)= \Tr\, \rho(\ln \rho-\ln\sigma)$
\cite[2.2,22(ii)]{Thirring}, since $S[\rho_{23}]-S[\rho_2]-S[\rho_3]=
- H(\rho_{23}, \rho_2\otimes\rho_3)$. By combining (\ref{65}) and
(\ref{66}) we have  thus  shown that
\begin{equation}
\liminf_{m\to\infty} \int_\Omega d\mu(\alpha)\, A_m(\alpha) \leq
\int_\Omega d\mu(\alpha)\, n^\alpha \big( S[\rho_{23}^\alpha]-
S[\rho_2^\alpha] - S[\rho_3^\alpha]\big).  
\end{equation}

It remains to show that
\begin{equation}\label{rts}
\liminf_{m\to\infty} \int_\Omega d\mu(\alpha)\, B_m(\alpha) 
\leq \int_\Omega d\mu(\alpha)\, n^\alpha S[\rho_3^\alpha],
\end{equation}
with 
\begin{equation}
B_m(\alpha) = \widehat n^\alpha S[\widehat \rho_3^\alpha] = S[\Tr_{12}\, 
\widehat\KK\widehat \K \widehat\rho_{123}] + \widehat n^\alpha \ln \widehat n^\alpha.
\end{equation}
Note that $\Tr_{12}\, \widehat\KK\widehat \K \widehat\rho_{123}\leq
P_3^{(m)} \rho_3 P_3^{(m)} \nl \rho_3$, since $\|\KK\K\|\leq 1$ and
hence $\|\widehat\KK\widehat \K\|\leq 1$.  By the same argument as
above, $\Tr_{12}\, \widehat\KK\widehat \K \widehat\rho_{123}$
converges to $\Tr_{12}\, \KK\K\rho_{123}$ in trace norm, and hence,
again by Theorem~A1 in \cite{LR},
\begin{equation}
\lim_{m\to\infty} B_m(\alpha) = n^\alpha S[\rho_3^\alpha] 
\end{equation} 
for each fixed $\alpha$. This gives pointwise convergence, but to show
convergence of the integral in (\ref{tre4}) we have to use the
dominated convergence theorem. We claim that $B_m(\alpha)$ is
uniformly bounded, independent of $m$ and $\alpha$. This is true since
$-x\ln x$ is monotone in $x$ for $0\leq x \leq 1/e$.  Since
$\Tr_{12}\, \widehat\KK\widehat \K \widehat\rho_{123}\nl \rho_3$, the
contribution to the entropy from eigenvalues less then $1/e$ is
bounded by the corresponding value for $\rho_3$.  Moreover, since
$\Tr\, \rho_3=1$, there are at most $2$ eigenvalues bigger than $1/e$.
This gives the claimed uniform bound.

By dominated convergence, we see that, for any subset $\Oo\subset \Omega$ 
with finite measure,
\begin{equation} 
\lim_{m\to\infty} \int_\Oo d\mu(\alpha)\, B_m(\alpha) = \int_\Oo d\mu(\alpha)\, 
n^\alpha S[\rho_3^\alpha] \leq  \int_\Omega d\mu(\alpha)\, 
n^\alpha S[\rho_3^\alpha]. \label{tog}
\end{equation}
Moreover, by concavity of $S[\rho]$, 
\begin{equation}\label{conc}
\int_{\Oo^c} d\mu(\alpha)\, B_m(\alpha) \leq  \left( \int_{\Oo^c} d\mu(\alpha)\, 
\widehat n^\alpha \right) S[\widehat\rho^c],
\end{equation}
with
\begin{equation}\label{rhoc}
\widehat \rho^c = \left[ \int_{\Oo^c} d\mu(\alpha)\, \widehat n^\alpha\right]^{-1} 
\Tr_{12} \int_{\Oo^c} d\mu(\alpha)\, \widehat \KK\widehat\K\widehat \rho_{123}. 
\end{equation}
Proceeding as in the proof of (\ref{ll}), we see that
\begin{equation}
\lim_{m\to\infty} 
\int_{\Oo^c} d\mu(\alpha)\, \widehat n^\alpha = \int_{\Oo^c} d\mu(\alpha)\, n^\alpha.
\end{equation}
Also $\Tr_{12} \int_{\Oo^c} d\mu(\alpha) \widehat
\KK\widehat\K\widehat \rho_{123} \to \Tr_{12} \int_{\Oo^c}
d\mu(\alpha) \KK\K\rho_{123}$ weakly, and thus in trace norm.  Since
again $\Tr_{12} \int_{\Oo^c} d\mu(\alpha) \widehat
\KK\widehat\K\widehat \rho_{123} \nl \rho_3$, the same argument as
above implies that the right side of (\ref{conc}) converges, in the
limit $m\to\infty$, to the corresponding expression without the
$\widehat{\phantom{x}}$'s. I.e.,
\begin{equation}\label{aaa} 
\limsup_{m\to\infty} \int_{\Oo^c}
d\mu(\alpha)\, B_m(\alpha) \leq \left( \int_{\Oo^c} d\mu(\alpha)\,
  n^\alpha \right) S[\rho^c], 
\end{equation} 
with $\rho^c$ given as in
(\ref{rhoc}), but with all the $\widehat{\phantom{x}}$ removed.

Now as $\Oo \to \Omega$, $\int_{\Oo^c} d\mu(\alpha)\, n^\alpha\to 0$.
Using again the dominance by $\rho_3$, which follows from the fact
that $\int_{\Oo^c}d\mu(\alpha)\, \KK\K \leq \id$, we see the right
side of (\ref{aaa}) goes to zero as $\Oo\to \Omega$. Together with
(\ref{tog}), this finishes the proof of (\ref{rts}), and hence the
proof of the theorem.

\section{Ruskai's proof of Corollary \ref{C2}}\label{ruskapp}

In an early version of this paper we had a weaker version of
Corollary~\ref{C2}, which read $S[\rho_{12}] - S[\rho_1] \leq S^{\rm
  cl}[\rho_{12}] - S^{\rm cl}[ \rho_{1}]$. In a private
correspondence, M.B.~Ruskai suggested the stronger version for the
case of (product) coherent states, and her suggestion motivated us to prove the
strengthened version of Corollary~\ref{C2} using our methods. The
following is a sketch of her proof.

1.) The map from a state to to its coherent state representation
is a completely-positive, trace-preserving map (CPT).

2.)  The {\it relative} entropy of two density matrices, 
$H(\rho, \mu) = \Tr \, \rho (\ln \rho -\ln \mu )$
is known to decrease under CPT maps. 
 
3.)  Apply 2.) to  $H(\rho_{12}, \rho_1 \otimes \rho_2)$ using the (product) coherent state
map, to obtain (\ref{weh}).

\end{document}